\documentclass[preprint,12pt]{elsarticle}

\usepackage{amssymb}
 \usepackage{amsthm}

\usepackage{hyperref}
\usepackage{xcolor}
\usepackage{graphicx}

\newcommand{\bgeqa}{\begin{eqnarray}}
\newcommand{\edeqa}{\end{eqnarray}}

\newcommand{\cmd}[1]{\texttt{#1}}

\definecolor{update}{rgb}{1,0,0} % red
\definecolor{update}{rgb}{0,0,0} % black
\definecolor{updateb}{rgb}{0,0,1} % green
\newcommand{\updated}[1]{{\color{update}#1}}

\journal{Computer Physics Communications}

\begin{document}

\begin{frontmatter}



\title{Gingred, a general grid generator for 2D edge plasma modeling}


\author{Olivier Izacard}
\ead{izacard@pppl.gov}
\author{Maxim V. Umansky}

%\cortext[corr]{izacard@llnl.gov}
\address{Lawrence Livermore National Laboratory, 7000 East Ave., Livermore, CA 94550}

\begin{abstract}
The current work reports the development of a new general grid generator called Gingred for arbitrary (e.g., number of X-points) 2D magnetic equilibria and plate geometries. A standardization of the construction of a grid is explained, the main command lines are detailed, and examples of snowflake grids are shown. The main advantages of Gingred are the tracking of the step-by-step progresses \updated{(from "skeleton" grids)} of grid generations by non-experts and new users, the ability to use an arbitrary magnetic configuration, and the more accurate numerical computation by bi-cubic interpolation of the magnetic flux rather than discrete cosine transform.
\end{abstract}


\end{frontmatter}


\section{Introduction}
\label{toc:Intro}
Understanding and predicting the physics at the edge of tokamaks is one of the keys for enhancing the efficiency of future fusion reactors. Transport fluid codes are used to simulate profiles that depend on the boundary conditions (core, shape of divertor plates and wall) in order to find good geometric configurations that reduce particle and heat fluxes on divertor plates. Possible solutions of the first wall protection against large particle and heat fluxes need to be assessed by simulating particle and power exhausts in the area in front of the divertor plates. To date, one of the most promising solutions to protect the first wall is to use an advanced divertor configuration. The original divertor (called standard divertor~\cite{bib_Divertor_S}) includes one X-point of the poloidal magnetic field where the primary separatrix is the magnetic flux surface which separates the closed (core) and the open (scrape-off layer, SOL) magnetic flux surfaces that hit the first walls at the divertor plates. The standard divertor has been extended to include additional secondary X-points (advanced divertor) close to the primary X-point. The presence of an additional secondary X-point can also reduce fluxes on divertor plates thanks to the parallel connection lengths, the flux expansions or the impurity radiations. These advanced divertor configurations have been called in the literature X~\cite{bib_Divertor_X}, super-X~\cite{bib_Divertor_SX}, snowflake-minus, snowflake-plus~\cite{bib_Divertor_SF} or clover-leaf~\cite{bib_Divertor_CL} (with two secondary X-points). These configurations are related to significant different physics phenomena but from the grid generation point of view, all of them are based on the unified idea of adding additional X-points. The secondary X-point can be relatively far away or close to the primary X-point in order to create a large area where the poloidal magnetic field is reduced (allowing different radiation and transport regimes in that area). Another possible effect is an increase of the poloidal flux expansion before the divertor plates (spreading the particle and power fluxes). Indeed, by focusing on the geometry we remark that the advanced divertor configurations include a secondary X-point that can be located far or close to the primary separatrix (i.e., X versus snowflake), inside or outside the SOL (i.e., snowflake-minus versus snowflake-plus), and before or after the plates (i.e., super-X versus snowflake). 
All these cases can be investigated if one is able to construct arbitrary grids. 
In this paper, the focus is on the advanced divertor geometry with only one additional X-point, called here by simplification the snowflake divertor (regardless of the leg length, the position of the secondary X-point with respect to the plate or the primary separatrix). It is straightforward to include more X-points following our bottom-up approach used in the code. \\
The first experiment with a snowflake divertor has been successfully performed on TCV~\cite{bib_TCV_SF}, followed by NSTX~\cite{bib_NSTX_SF} and DIII-D~\cite{bib_DIIID_SF}. It turns out that there is a large number of free parameters (e.g., position of the secondary X-point, geometry of the plates, distance between the X-points, distance between the X-points and the plates). Therefore, we must run many numerical simulations of the divertor plasmas in order to find the best configurations that reduce the fluxes at the plates without reducing the core performances. Continuous or particle based kinetic numerical codes (e.g., COGENT~\cite{bib_COGENT}, XGC1~\cite{bib_XGC1}) are under development for simulating divertor plasmas and have a large CPU-consumption, but fluid codes with smaller CPU-consumption have been developed and validated against experiments. Currently, the state-of-the-art transport fluid codes are SOLPS-Iter~\cite{bib_SOLPS-Iter} (recent choice of Iter Organisation, not benchmarked yet), EMC3~\cite{bib_EMC3} (3D but no charge-state resolved impurity model) and UEDGE~\cite{bib_UEDGE} (2D with a possible charge-state resolved impurity model). All fluid codes require a grid, i.e., the mesh on which the fluid equations are computed. Moreover, other fluid codes that focus on turbulence (e.g., BOUT++~\cite{bib_BOUT++}) use very similar grids. \updated{Even the next generation of fluid codes that will include non-Maxwellian kinetic effects~\cite{Izacard_2016_INMDF,Izacard_2017_INMDF} would be able to use those grids.} It is then required to be able to manage the creation of arbitrary grids from arbitrary magnetic configuration for arbitrary codes. We notice that UEDGE has the capability to internally generate the grid which but that has been initially developed for single- and double-null magnetic geometries. 
%\updated{ Indeed only a few examples of snowflake grids have been obtained to prove the capability to create those grids (more developments are still under development~\cite{Rensink_2016}). 
%More developments of snowflake grids construction by the UEDGE internal grid generator are in progress~\cite{Rensink_2016}. 
\updated{The spatial indexing scheme that allows the various plasma edge configurations to be represented on a logically rectangular mesh was developed previously by many groups~\cite{%bib_Khosla_1974,
bib_Braams_1983,
bib_Subba_2004,
bib_McCourt_2012} for single- and double-null magnetic configurations. According to our knowledge, the first snowflake logical maps for some snowflake configurations have been developed by Rensink~\cite{bib_Rensink_2017} and is illustrated in~\ref{toc.Appendix}. \footnote{\updated{Following those ideas of logical maps, a universal version of the code Gingred is under development for the construction of arbitrary logical maps with an arbitrary number of X-points and plates.}} }
However, the UEDGE internal grid generator is not adapted for the unified step-by-step bottom-up development of the grid that is crucial for debugging. There exist other grid generators: BOUT++ uses Hypnotoad~\cite{bib_BOUT++} and SOLPS-Iter uses CARRE~\cite{bib_CARRE} which both have a graphical user interface (IDL widgets for Hypnotoad and the DG code for CARRE), but to date Hypnotoad and CARRE do not include the step-by-step construction of the grids and do not manage snowflake grids. For all of these reasons, our interest is to develop a new universal code toward the development of arbitrary grids, independent of all fluid code, and easily modifiable by new users following a robust and simple (i.e., small number of command lines) step-by-step workflow. \\
We develop here the IDL-based Gingred code for grid generation that is robust, universal, sequential and easily manageable by new users with respect to the main existing grid generators thanks to new algorithms and the bottom-up approach. Indeed, Gingred works with a few simple command lines, such as splitting a cell of a mesh in 2, which allows to construct arbitrary grids (i.e., regardless of the leg length or the flux expansion). The name Gingred has been chosen by inverting the letters ``i" and ``e" from the common ``GenGrid" or ``GridGen" names. The execution of Gingred from the integrated framework OMFIT~\cite{bib_Meneghini_2013_PFR_8,bib_Meneghini_2015_NF_IAEA} is in progress. The work on Gingred was initiated in 2012 to simplify and create an interactive grid generator for the construction of snowflake grids for UEDGE simulations~\cite{Rognlien_2014_APS,Izacard_2016_APS}. One of the advantages of Gingred is the capability to see each step of the grid construction from automatic plots. This {feature helps with the creation of arbitrary grids by new users and with the debugging and development of additional capabilities of the code by developers. Gingred is accessible via the Version Control with Subversion~\cite{bib_SVN} (called svn). Svn is similar to Github~\cite{bib_Github} for sharing projects between developers and managing different versions of a code. \\
This article is organized as follow. Sec.~\ref{toc.Grids} contains the description of several magnetic configurations that require a different logical grid topology. The latest version of the code manages the following grids: no X-point, a single-null, a double-null, two snowflake-minus, and two snowflake-plus magnetic configurations. However, the extension to new magnetic geometries would be straightforward following the existing scripts and the details shown here. {Sec.~\ref{toc.Gingred} provides a description of the code and the main command lines, and shows examples of snowflake grid generation. Sec.~\ref{toc.Future} deals with possible future developments of Gingred.

\section{Available grids}
\label{toc.Grids}
This section summarizes the different grids and notations used for different magnetic equilibria. It turns out that the different magnetic geometries related to the grids shown in~\ref{toc.Appendix} can be universalized by using a minimal set of parameters. We show values of this minimal set of parameters for the main magnetic equilibria. Initially developed in the literature~\cite{bib_CARRE,bib_UEDGE,bib_Rensink_2017}, each grid configuration is associated with a 2D map of indices of cells (called the 2D logic map). Here we chose to standardize these 2D logic maps by using parameters in order to easily switch between various grid configurations. The position of the X-point(s) and the plates are determined by the minimal set of parameters such as the arrays \cmd{ixpt1}/\cmd{ixpt2} for the inner/outer index position of X-point(s), and \cmd{ixlb}/\cmd{ixrb} for the index of left/right boundary (plates) positions. Only for the left boundary \cmd{ixlb} the indices correspond with the cell of the poloidal position of the left plates in the 2D logic map, but for all the other arrays the indices correspond with the position of the previous poloidal cell. This choice of indices follows the code UEDGE because Gingred was initially developed for the grid generation of advanced divertor configurations for UEDGE simulations. In~\ref{toc.Appendix}, we show the correspondence between the main magnetic geometries and their 2D logic map of cells (also called minimal grid of patches). Once we understand how to switch between each magnetic geometries with different number of plates and different {X-point(s) and plates positions, it becomes trivial to add more X-points, but the number of different 2D logic map of cells quickly increases. Indeed, we have $1$ main configuration with one X-point, $4$ main configurations with two X-points, and $16$ main configurations with three X-points (without taking into account special cases where two X-points are on the same magnetic flux surface). \\
We explicitly show the minimal 2D logic map of cells in~\ref{toc.Appendix} for the main grid configurations. Each map is different due to the relative position of plates and X-point(s). The bottom-up approach of Gingred starts by creating these minimal 2D logic maps \updated{(i.e., the "skeleton" grids)}, then the refinement of the grid configuration can be done independently of the 2D logic map under consideration. \\
Moreover, it is straightforward to expand the 2D logic map of cells to include more X-points such as one upper X-point with a lower snowflake geometry or the cloverleaf (3 X-points) magnetic geometry~\cite{bib_Divertor_CL}. The generalization of the code with an arbitrary number of X-points and plates is under development.

\section{Gingred grid generator}
\label{toc.Gingred}
This section reports some details of the Gingred grid generator. The code, written in IDL, manages the construction of all magnetic configurations detailed in Sec.~\ref{toc.Grids} {as well as the case with no X-point relevant to the edge of reversed field configurations (FRC).

%\subsection{Pre-requirements}
%In order to use Gringred, one can import the code using Subversion with the command:\\
%%\begin{lstlisting}[language=Python]
%%svn co svn+ssh://username@portal-auth.nersc.gov/project/projectdirs/edge_mdl/gingred
%%\end{lstlisting}
%\cmd{svn co svn+ssh://USER@portal-auth.nersc.gov/project/projectdirs/\\edge\_mdl/gingred gingred/} \\
%where \cmd{USER} is your NERSC account and \cmd{gingred/} is the destination directory. Documentation on SVN can be found online~\cite{bib_SVN}. %\\
In order to use Gringred, one can import the code using Subversion \cmd{svn} command line. 
All other IDL functions are included in the SVN package.

\subsection{List of available commands}
The main commands necessary to run Gingred and to construct a grid (a snowflake-minus SF45 grid is taken as example, see~\ref{toc.Appendix}) are listed by order of appearance for the grid generation. The options are shown with brackets \cmd{[option=value]} or \cmd{[/option]} for a flag. \\
- \cmd{read\_efit[, /plot]} \\
This command line imports the magnetic equilibrium from the geqdsk files (note that geqdsk can be a link to a file).\\
- \cmd{mapping[, importfile="gp.SF45", gridparams=gp, conf="SF45", \\/manual, /ascii]} \\
This command line setup the boundaries of the grid. The user interface is used by the manual flag \cmd{/manual} for the magnetic configuration set by \cmd{conf="SF45"} (the available geometries are FRC, SN, DN, SF15, SF45, SF75 or SF95), whereas the automatic mode can be used if one set \cmd{importfile=} \cmd{"gp.SF45"[, /ascii]} where gp.SF45 is an idl.save file (or an ascii file if the flag \cmd{/ascii} is used) file. Similarly, the automatic mode can be used by importing \cmd{gridparams=gp} where \cmd{gp} is a local IDL structured variable defining the grid parameters. Examples of those IDL and ascii files are given and they can be created after the execution of the interactive mode. For the interactive mode (with the flag \cmd{/manual}), the user has to select consecutively point positions close to: the magnetic field center, the primary X-point, and the secondary X-point (for non single-null grids). Then the user has to choose the flux surface boundaries (represented by different colors in the 2D logic maps) of: the primary SOL (yellow), the core (cyan), the primary private flux region (red), and the following additional choices for non single-null grids: the secondary private flux region (purple) and the secondary SOL (grey). It is also possible to fix the boundary fluxes, and the plate boundaries from their values instead of using the user interface with the command \cmd{grid\_params, gp, name='psi\_core', value=Get\_PsiN\_Inv(0.9), /set\_var}, where the function \cmd{Get\_PsiN\_Inv} returns the non-normalized poloidal flux from the normalized value given as input. For example, the most used variables one can set are \cmd{psi\_core}, \cmd{psi\_sol1}, \cmd{psi\_sol2}, \cmd{psi\_pf1} and \cmd{psi\_pf2} flux surfaces, and \cmd{magx}, \cmd{core}, \cmd{xpt1}, \cmd{xpt2}, \cmd{istrike1}, \cmd{istrike2}, \cmd{ostrike1} and \cmd{ostrike2} points, at least for a snowflake magnetic configuration with two X-points. For the type of the variables: fluxes are double variables, and points are structures with 5 quantities (\cmd{r} and \cmd{z} the position coordinates, \cmd{psi} the magnetic flux, \cmd{s} a length coordinate internally used by the code, and \cmd{type} a logical variable equal to 1 if the point is a X-point). \\ 
- \cmd{grid\_params, gp[, importfile="gp.SF45", exportfile="gp.SF45.new", /reset, /ascii]} \\
This command line creates a local copy of the variable \cmd{gp} accessible by the user that contains some parameters for the grid such as the position of the magnetic axis (i.e., \cmd{gp.magx}), the position of a point delimiting the core boundary (i.e., \cmd{gp.core}), or the position of the X-point(s) (i.e., \cmd{gp.xpt1} and \cmd{gp.xpt2}). The list of available parameters can be shown with the command \cmd{help, gp}. These parameters can be imported from or exported to a IDL (respectively ascii) file by using the options such as \cmd{importfile} or \cmd{exportfile} (respectively by using the function \cmd{read\_gp} and \cmd{write\_gp} when the flag \cmd{/ascii} is used). \\
- \cmd{show\_domain[, /xpt, /plates, yr=[-1.8,-1.2], xr=[.3,.9]]} \\
This command line is used to plot boundary flux surfaces, X-point(s) structures (flag \cmd{/xpt}), and plate boundaries (flag \cmd{/plates}). A zoom can be managed by the option \cmd{xr} and \cmd{yr}.\\
- \cmd{gridcells\_all, g1[, conf="SF45"]} \\
This command line constructs the initial grids described in Sec.~\ref{toc.Grids} and shown in~\ref{toc.Appendix}. The variable \cmd{g1} is returned as an array of cells of dimension $(nx,ny)$. Each cell is a structure including the following points: \cmd{ptNW, ptSW, ptNE, ptSE, ptN, ptS, ptW, ptE} and the variable \cmd{guard} that defines if the cell is a guard cell used for plate positions. The first 4 points are summits of the cell, and the last 4 points are midpoints used to split the cell. Finally, each point is a structure detailed below (see the command line defining \cmd{ptN}). \\
- \cmd{fixmidpt, g1, g2[, /advanced]} \\
This command line fixes the poloidal position of the midpoints (i.e., \cmd{ptN} and \cmd{ptS}) of each cells (see more details with the command line \cmd{refgridp}). The output grid is \cmd{g2}. The flag \cmd{/advanced} can be used to fix midpoints based on the midpoint that is on the primary separatrix. \\
- \cmd{doublep, g1, g2[, /fix]} \\
This command line is used to double the number of poloidal cells from g1 and creates g2. The flag \cmd{/fix} corrects the poloidal midpoints by calling \cmd{fixmidpt}. \\
- \cmd{doubler, g1, g2[, /fix]} \\
This command line is used to double the number of radial cells from g1 and creates g2. \\
- \cmd{refgridp, g1, g2, i0=3[, yr=[-1.8,-1.2], xr=[0.3,0.9]]} \\
This command line splits the poloidal cells of index \cmd{i0=}$3$ in two. In other words, a new poloidal line appears in the cells \cmd{i0=}$3$. The new poloidal line follows the gradient of the poloidal magnetic flux. We remark that the midpoints of each cells \cmd{i0=3} are defined but the new poloidal line almost never coincide with these midpoints. This issue is the reason of the developed \cmd{fixmidpt} script which construct the new poloidal lines as close as all cells midpoints, even through it follows the gradient of the poloidal magnetic flux. The output grid is \cmd{g2}. \\
- \cmd{refgridr, g1, g2, j0=1[, yr=[-1.8,-1.2], xr=[0.3,0.9]]} \\
This command line splits the radial cells of index \cmd{j0=}$1$ in two. The output grid is \cmd{g2}. \\
- \cmd{ptN=\{r:0d0,z:0d0,psi:0d0,type:0,s:0d0\}} \\
The users need to set list of points to define each plate. This structure contains the radial \cmd{r} and vertical \cmd{z} coordinates, the poloidal flux \cmd{psi}, the logical \cmd{type} equal to 1 if the point is a X-point, and a curved integrated length \cmd{s}. \\
- \cmd{p1a=ptN \& p1a.r=0.2 \& p1a.z=-1 \& p1a.psi=Get\_psi(p1a.r,p1a.z)} \\
This is an example to modify the point \cmd{p1a}. The function \cmd{Get\_psi(r,z)} is defined to return the value of the magnetic flux surface at the coordinate $(r,z)$. Note that the functions  \cmd{Get\_psiN(psi)} and \cmd{Get\_psiN\_Inv(psiN)} also exist and return respectively the normalized poloidal flux (\cmd{psiN}) and its non-normalized value (\cmd{psi}). \\
- \cmd{pltarr=list([p1a,p1b],[p2a,p2b],[p3a,p3b],[p4a,p4b])} \\
To setup $4$ plates with only $2$ points each (make sure that each plate intersects the domain). We can have as many segments as wanted to define each plate because Gingred detects which segment of the plate intersects a specific magnetic flux surface during the non-orthogonal grid modification by plate constraints. By default the code keeps the ratio of the poloidal lengths from the orthogonal grid. This option is ideal when the original orthogonal boundary curve is close to the plate geometry. However, for more complex plate geometry, the option \cmd{/equal} can be used to have equal poloidal lengths of the cells in the divertor legs. Even if the later option is not often used, we remark that it induces larger angles between the edges of adjacent cells (see Figs.~\ref{fig.Gingred.SF45}(c) and~\ref{fig.Gingred.SF75}(c)) that can facilitate numerical convergence due to the area around the primary X-point (see BOUT++ grids generated by Hypnotoad). \\
- \cmd{plates\_add, g1, g2[, /equal, /fastrun\updated{, ixortho=ixortho,\\ wplt=[1,0,0,0]}]} \\
This command line reconstructs non-orthogonal grids for each legs with \cmd{/equal} or proportional (by default) poloidal lengths, and \cmd{/fastrun} is an option that can be used when the longest poloidal lengths of the legs are less than 1 meter. The output grid is \cmd{g2}. The script uses the parameters saved in the gp variable such as the poloidal position of X-points and plates in the 2D logic map of indices. \updated{The option given by the array \cmd{wplt} specify which boundary is going to be constrained by the plate geometry given by the list \cmd{pltarr}. By default \cmd{wplt} is an array of ones in order to constrain each plate geometry. The matrix \cmd{ixortho} of dimension $(nplt,ny)$\footnote{\updated{Here, $nplt$ is the number of plates and $ny$ the number of radial cells.}} is setup by default to the indices $ixpt1$ and $ixpt2$ of the poloidal positions of the X-point(s) in order to constrain all cells in each leg. With this option, users can constraint only a part of the private flux region, and more cells in the SOL above the X-point(s) positions.}\\
- \cmd{plates\_plot} \\
To plot the plates defined by the list of arrays \cmd{pltarr} \\
- \cmd{generate\_header, g1, hd1} \\
The output variable \cmd{hd1} is the header needed for the creation of the ``gridue" file. \\
- \cmd{export\_grid, g1, h=hd1, /save} \\
It exports the grid \cmd{g1} into the file ``gridue". The guard cells are constructed before to export the grid.\\
All of these command lines can be used for the generation of grids with different sizes and refinements. Examples of grid generation with plate geometry constraints are shown in Secs.~\ref{toc.Gingred.example1} and~\ref{toc.Gingred.example2}.\\

\subsection{Algorithms used in Gingred}
\label{toc.Gingred.algo}
The approach used in this work is to produce a minimal grid of patches called the ``skeleton" grids (as described in Sec.~\ref{toc.Grids} and shown in \ref{toc.Appendix}), consistent with a given magnetic equilibrium configuration. Then, each patch can be divided poloidally and radially until a desired grid is built. Moreover, there are other advanced algorithms used by Gingred. In comparison to the literature where the discrete cosine transform is used for the interpolation of the poloidal magnetic flux, Gingred has the capability to use a bi-cubic interpolation. We believe that the bi-cubic interpolation is more robust, especially for snowflake magnetic configuration where a large area of small poloidal magnetic field is present as well as a low refinement of the magnetic equilibrium geqdsk input file (usually on a 64x64 mesh). Based on this interpolated magnetic field, the code can use ODE integrators such as LSODE and Runge-Kutta (RK4) for the computation of radial and poloidal lines. The integrator LSODE is more robust than RK4, when we use the default error tolerance parameters. Radial lines are the curves following perpendicular directions of the flux surfaces defined by iso-contours of the poloidal magnetic flux, and poloidal lines are the curves following the iso-contours of the poloidal magnetic flux. From these capabilities, the code manages the construction of the orthogonal grid and arbitrary refinements of the grid decided by the user. For the non-orthogonal capability of the code, the length of the poloidal curves following iso-contours of the flux surfaces are used to determine the position of the cells. Finally, for the refinement of the number of radial cells, the code uses the default IDL root finding in order to find a value of the flux surface along a radial line. All of these algorithm are robust and there is no bottleneck that significantly reduces the computation time.

\subsection{Example of a SF45 grid generation}
\label{toc.Gingred.example1}
Here is a list of command lines used for the generation of the SF45 grid for a simulated magnetic equilibrium on NSTX-U~\cite{Soukhanovskii_IEEE_2016}. The link of the geqdsk magnetic equilibrium file (where $\$DIR$ refers to \cmd{/}\textit{your\_dir}\cmd{/gingred/data/SF45}) is setup by: \\
%\cmd{> ln -s $\$DIR$/a135111.00600.Njw\_08wf-SFDm3 aeqdsk} \\
\cmd{> ln -s $\$DIR$/geqdsk neqdsk} \\
Then, from the initial $13 \times 5$ (with guard cells used for the plates) grid we started by doubling three times the number of radial cells (\cmd{doubler}) and one time the number of poloidal cells (\cmd{doublep}) and we obtain a $22 \times 26$ grid (with guard cells). Then, the poloidal refinement of the grid is done by executing the \cmd{refgridp} command where
\bgeqa
\displaystyle
i0 \in \{ 20, 21, 19, 19, 19, 18, 14, 13, 13, 13, 10, 11, 12, 13, 14, 15, 3, 3, 3, 3, 3, 2, 1 \}.
\edeqa
Finally, the radial refinement of the grid is done by executing the \cmd{refgridr} command where $j0 \in \{ 17,17,8,9,10,11 \}$. The constraint of plate geometries is possible with: \\
\cmd{
IDL> pltarr=list([p1a,p1b],[p2a,p2b],[p3a,p3b],[p4a,p4b]) \\
%IDL> gp.ixpt1 => [5,27,38,38] \& ixplt = [0,33,35,43] \\
IDL> plates{\_}add, gf4f29f, gf4f29fp, yr=[-1.8,-1.2], xr=[.3,.9] \\
IDL> show{\_}grid, gf4f29fp, yr=[-1.8,-1.2], xr=[.3,.9] \& plates{\_}plot \\
IDL> generate{\_}header, gf4f29fp, hdf4f29fp \\
IDL> grid{\_}export, gf4f29fp, h=hdf4f29fp, /save \\
%IDL> \$mv gridue gridue.NSTXU.135111.SF75.43x30.no \\
}
We remark that the arrays \cmd{gp.ixpt1}, \cmd{gp.ixpt2}, \cmd{gp.ixlb} and \cmd{gp.ixrb} are defining the leg indices where the non-orthogonal grid is computed and the flag \cmd{/equal} can be added in the command line \cmd{plates{\_}add} in order to use equal poloidal cell lengths. The flag \cmd{/fastrun} can be used to avoid additional search of the smallest poloidal distance in case the first direction gives a distance smaller than $1$ m (i.e., one do not should use this option when legs are too close to a poloidal magnetic O-point). The function \cmd{plates{\_}plot} shows the plates defined by \cmd{pltarr}. Fig.~\ref{fig.Gingred.SF45} shows (a) the grid \cmd{gf4f29f} without plate constraints, (b) the grid \cmd{gf4f29fp} with plate constraints and (c) the grid \cmd{gf4f29fp2} with plate constraints and the flag \cmd{/equal}.

\begin{figure}[!htbp]
\centerline{
$\begin{array}{cc}
\includegraphics[width=82mm]{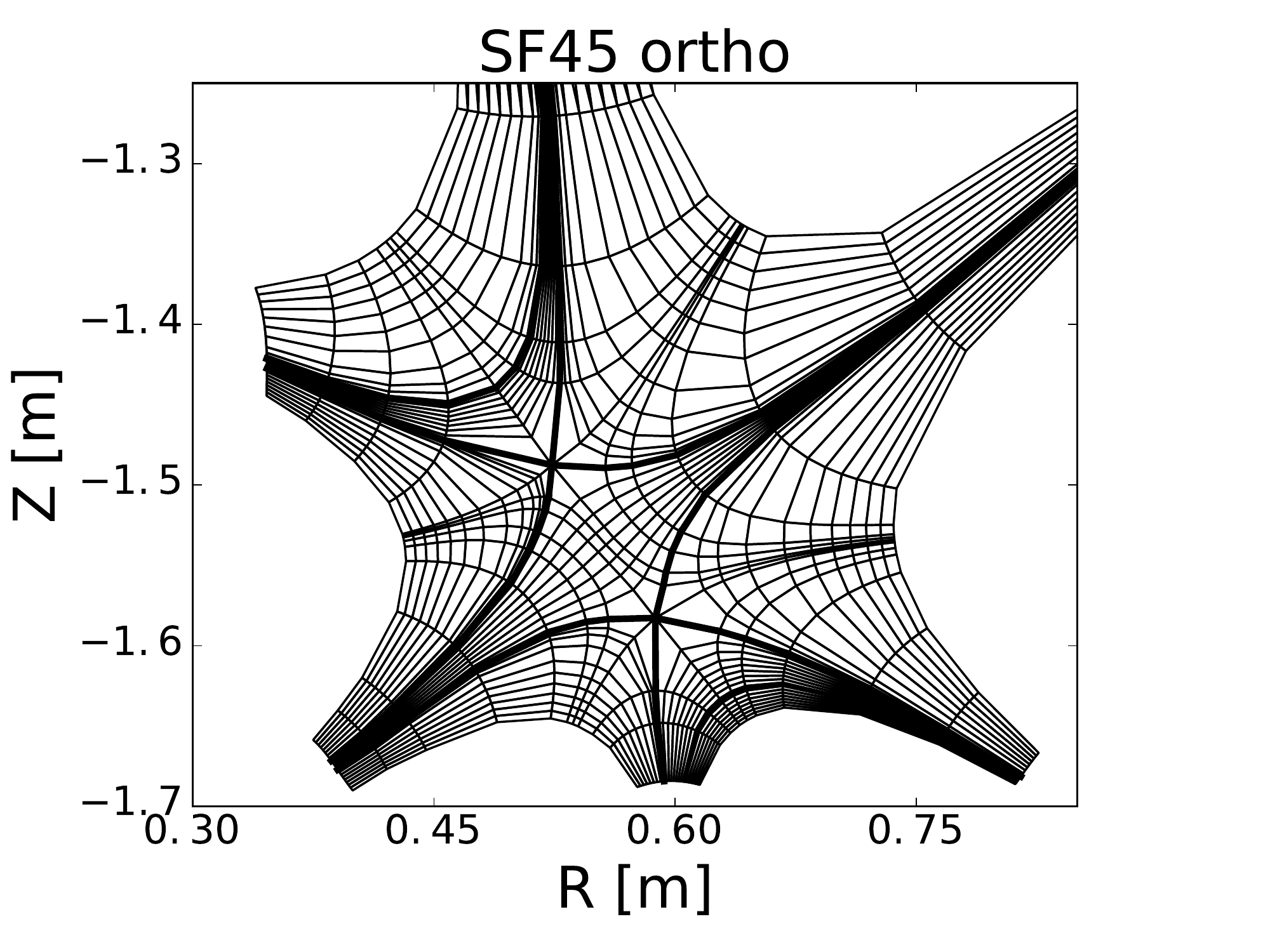}
&
\includegraphics[width=82mm]{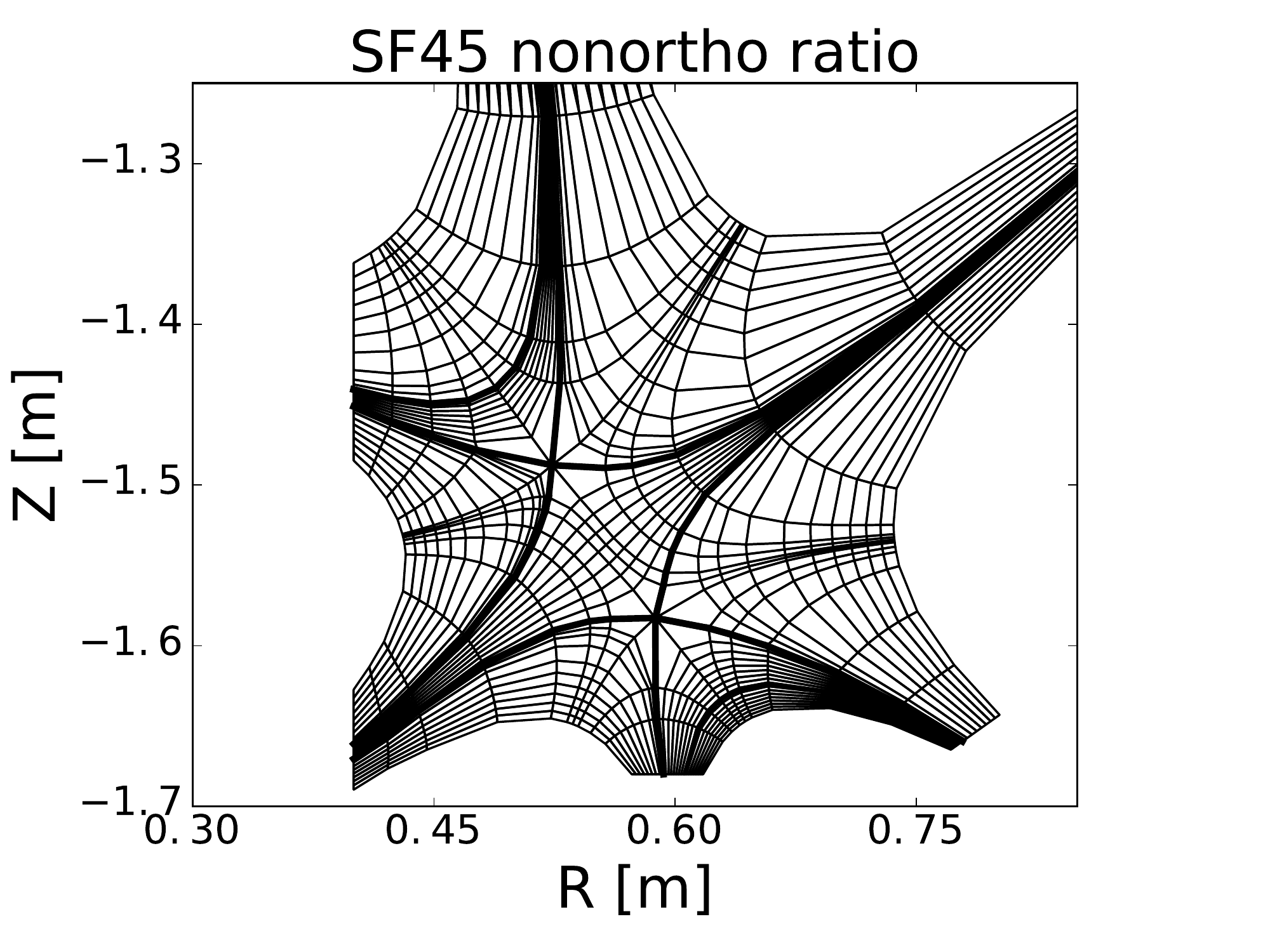}
\\
\label{fig.Gingred.SF45.o} \textrm{(a) Orthogonal SF45}
&
\label{fig.Gingred.SF45.nor} \textrm{(b) Nonorthogonal SF45}
\\
 & \textrm{with plate constraints.} \\
\includegraphics[width=82mm]{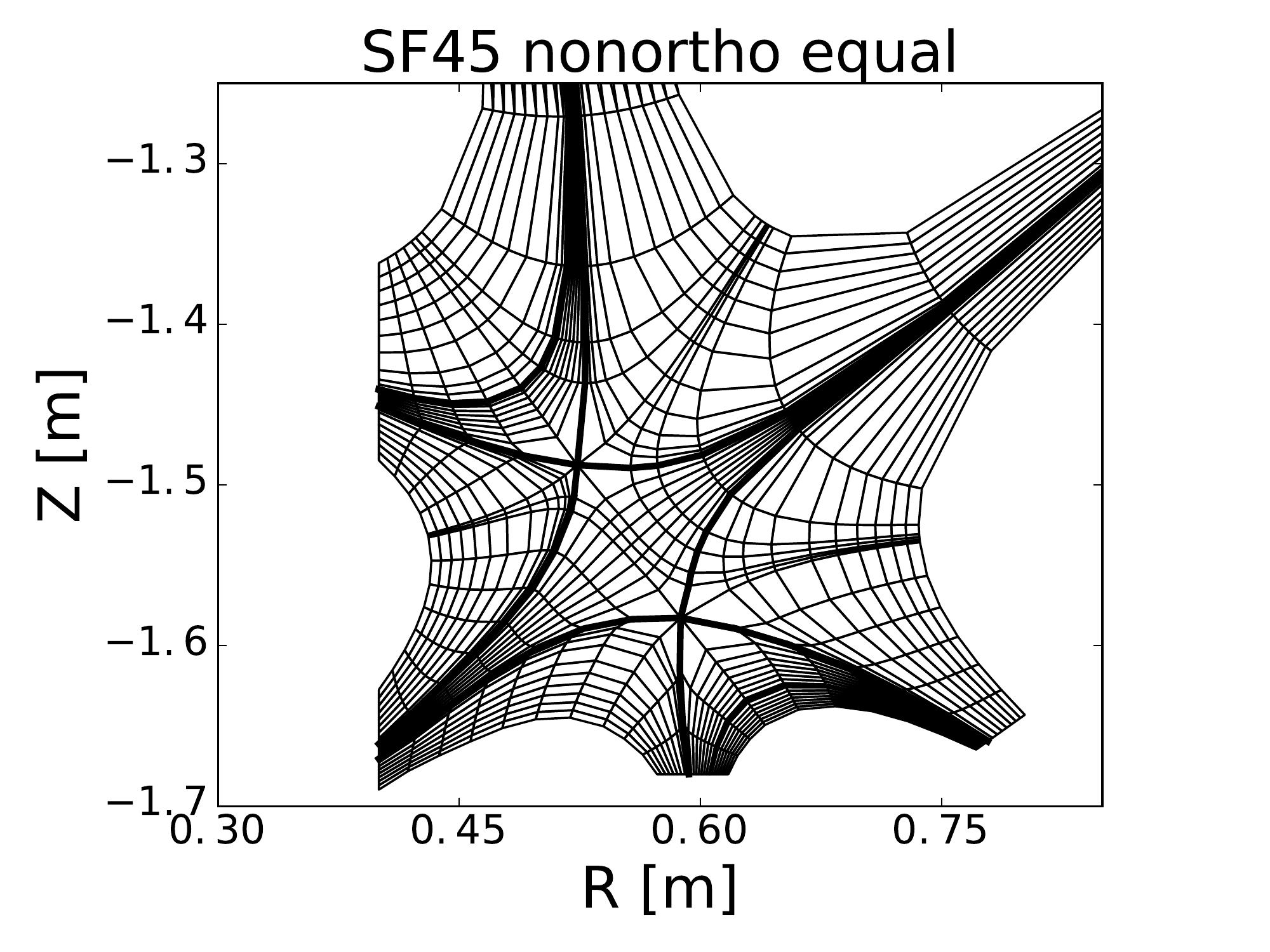}
\\
\label{fig.Gingred.SF45.noe} \textrm{(c) Nonorthogonal SF45}
\\
 \textrm{with "equal" plate constraints.}
\end{array}
$}
\caption{Examples of snowflake-minus grids without (a), and with (b) and (c) plate constraints. Figs. (b) and (c) are obtained by keeping the ratio of poloidal lenthgs (b) or by using equal poloidal lengths (c).}
\label{fig.Gingred.SF45}
\end{figure}

\subsection{Example of a SF75 grid generation}
\label{toc.Gingred.example2}
Here is a list of command used for the generation of the SF75 grid for a simulated magnetic equilibrium on NSTX-U~\cite{Soukhanovskii_IEEE_2016}. The link of the geqdsk magnetic equilibrium file (where $\$DIR$ refers to \cmd{/}\textit{your\_dir}\cmd{/gingred/data/SF75}) are setup by: \\
%\cmd{> ln -s $\$DIR$/a135111.00600.mod-SFp1{\_}oi aeqdsk} \\
%\cmd{> ln -s $\$DIR$/g135111.00600.mod-sfdp neqdsk} \\
\cmd{> ln -s $\$DIR$/neqdsk neqdsk} \\
Then, from the initial $13 \times 5$ grid we started by doubling three times the number of radial cells and one time the number of poloidal cells and we obtain a $22 \times 26$ grid. Then, the poloidal refinement of the grid is done by executing the \cmd{refgridp} command where
\bgeqa
\displaystyle
i0 \in \{ 20, 19, 19, 18, 19, 14, 13, 13, 12, 11, 10, 11, 12, 13, 14, 3, 3, 3, 3, 3, 2, 3, 1 \}.
\edeqa
Finally, the radial refinement of the grid is done by executing the \cmd{refgridr} command where $j0 \in \{ 17,17,16,17,18,8 \}$. The constraints of plate geometries are possible, similar to the previous example. %\\
%\cmd{
%IDL> pltarr=list([pt1a,pt1b],[pt2a,pt2b],[pt3a,pt3b],[pt4a,pt4b]) \\
%IDL> ixpt = [5,27,38,38] \& ixplt = [0,32,34,43] \\
%IDL> plates{\_}add, gf4f29f, gf4f29fp, yr=[-1.8,-1.2], xr=[.3,.9] \\
%IDL> show{\_}grid, gf4f29fp, yr=[-1.75,-1.2], xr=[.3,.85] \& plates{\_}plot \\
%IDL> generate{\_}header, gf4f29fp, hdf4f29fp \\
%IDL> grid{\_}export, gf4f29fp, h=hdf4f29fp, /save \\
%IDL> \$mv gridue gridue.NSTXU.135111.SF75.43x30.no \\
%}
Fig.~\ref{fig.Gingred.SF75} shows (a) the grid \cmd{gf4f29f} without plate constraints, (b) the grid \cmd{gf4f29fp} with plate constraints and (c) the grid \cmd{gf4f29fp2} with plate constraints and the flag \cmd{/equal}.\\
\begin{figure}[!htbp]
\centerline{
$\begin{array}{cc}
\includegraphics[width=82mm]{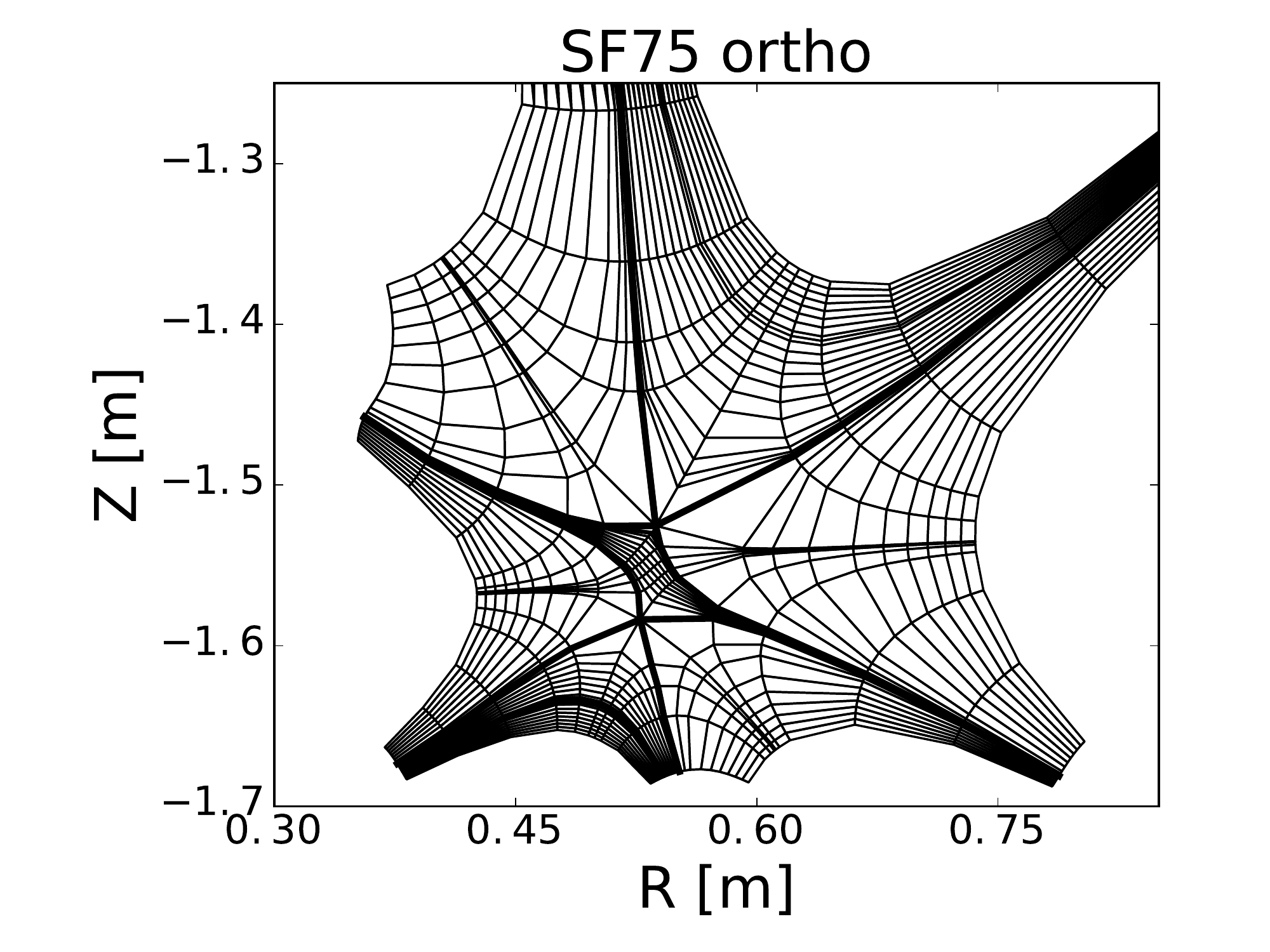}
&
\includegraphics[width=82mm]{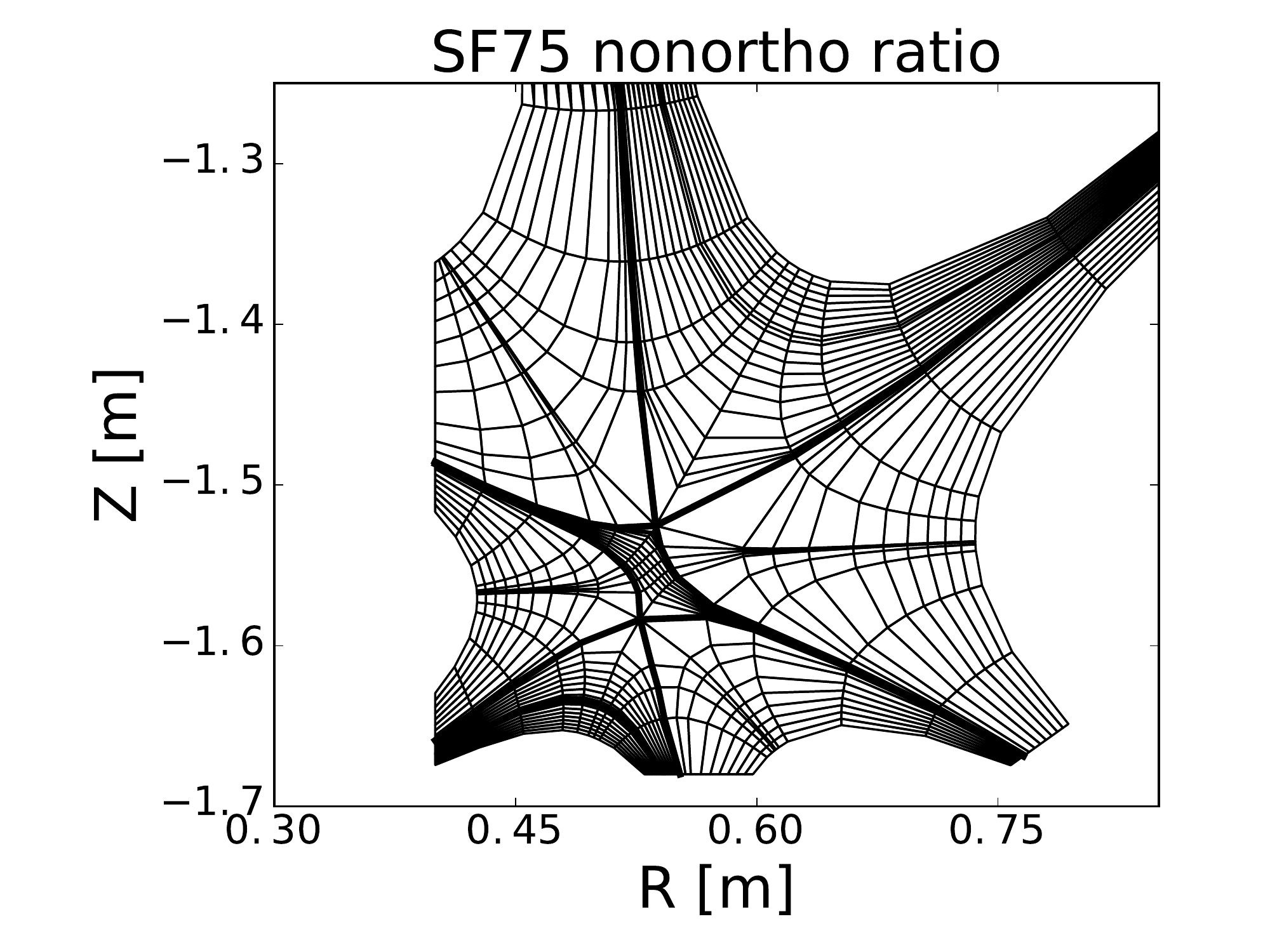}
\\
\label{fig.Gingred.SF75.o} \textrm{(a) Orthogonal SF75}
&
\label{fig.Gingred.SF75.nor} \textrm{(b) Nonorthogonal SF75}
\\
 & \textrm{with plate constraints.} \\
\includegraphics[width=82mm]{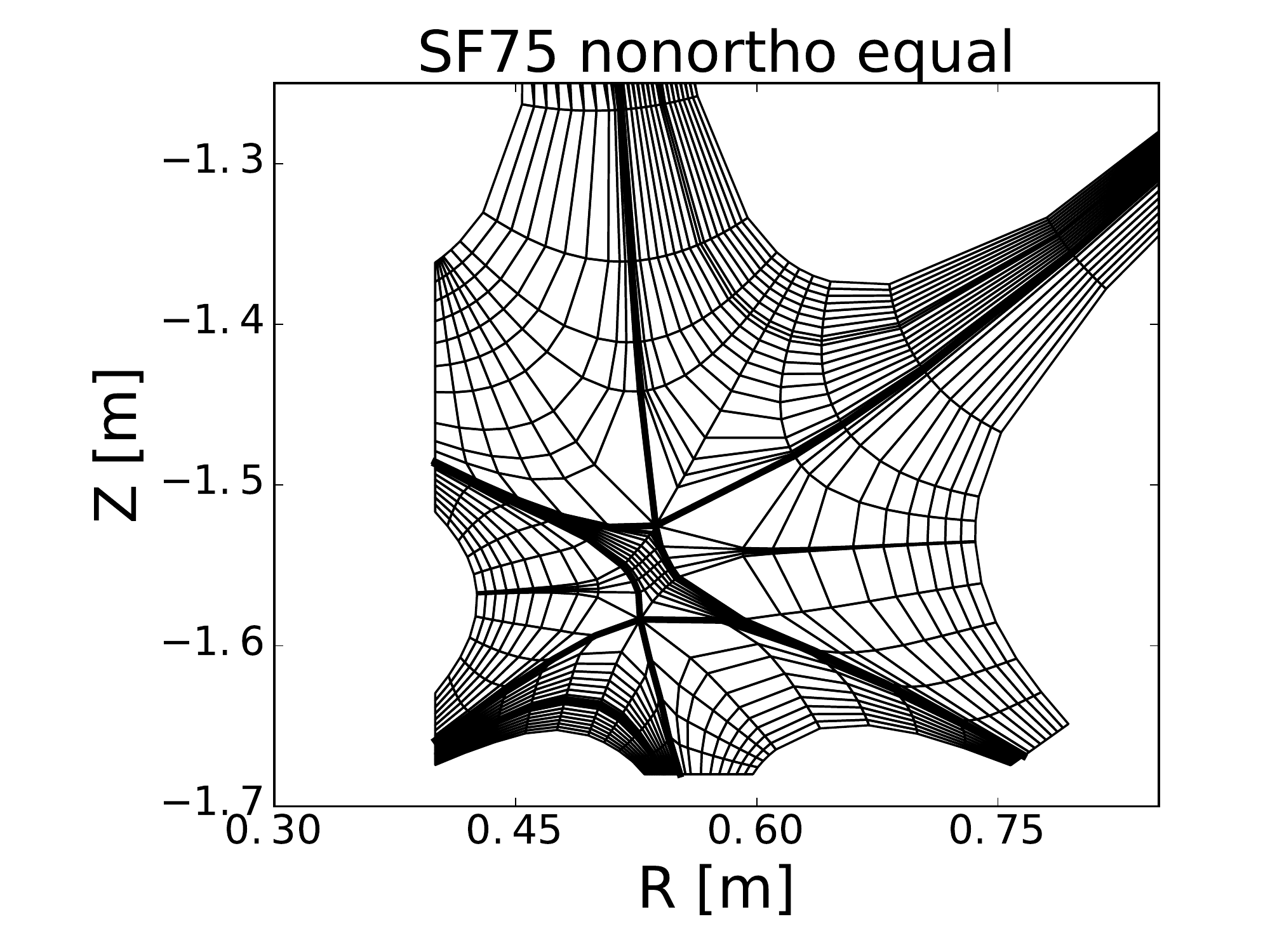}
\\
\label{fig.Gingred.SF75.noe} \textrm{(c) Nonorthogonal SF75} \\
 \textrm{with "equal" plate constraints.}
\end{array}
$}
\caption{Examples of snowflake-plus grids without (a), and with (b) and (c) plate constraints. Figs. (b) and (c) are obtained by keeping the ratio of poloidal lengths (b) or by using equal poloidal lengths (c).}
\label{fig.Gingred.SF75}
\end{figure}

In addition to these examples, the code contains more detailed documentation as well as examples of automatic generation of grids from a given input option file. The input option file, an ascii or an IDL structure, can be used in by the \cmd{mapping} function (without the flag \cmd{/manual}) in order to automatically build the initial grid of patches. To date, there is no support for the full automatic generation of the input option file (find X-point, detection of magnetic configuration, standard position of plates and boundary flux surfaces). Instead, they are created by a interactive run of the function \cmd{mapping} with the flag \cmd{/manual}.

\updated{
\section{Comparison with state of the art grid generators}
\label{toc.OtherGridGen}
The new grid generator Gingred is unique by comparing existing grid generators. First, according to our knowledge there is no existing grid generator that is capable of constructing a grid for snowflake magnetic configurations except the internal grid generator of the code UEDGE~\cite{bib_Rensink_2017}. The reason is mainly due to the unique extension, for the first time in Ref.~\cite{bib_Rensink_2017}, of the logic maps from single-null and double-null to the new classes of logic maps for snowflake configurations (see~\ref{toc.Appendix}). However, the internal grid generator of UEDGE is not accessible without UEDGE and it has not been tested yet with new users. Gingred has been successfully tested~\cite{Izacard_2016_APS} by new users~\cite{Kaptanoglu_2017} for reversed field configurations, and different grids have been created for multiple machines (e.g., NSTX, NSTX-U, DIII-D and ADX) with complex plate geometries such as the Small Angle Slot divertor installed at DIII-D~\cite{Guo_2017}. In other words, Gingred is the first numerical application of the new logic maps for snowflake configurations accessible for multiple codes. However, there are fundamental differences between the internal grid generator of UEDGE and Gingred with respect to the snowflake grid generation. As examples, (i) Gingred is focused on a bottom-up approach at the level of the user interaction that ease significantly the progresses of the grid generation even for non-experienced users in Gingred
% (see Refs.~\cite{McGreivy_2016_APS}  and~\cite{Kaptanoglu_2017})
 (see Ref.~\cite{Kaptanoglu_2017})
, (ii) the code includes multiple options for debugging, plotting, or changing the accuracy at each step of the workflow, (iii) robust automatic scripts have been developed and are included in the code as practical examples, (iv) Gingred can be used to generate grids for arbitrary kinetic or fluid codes, and (v) the new bi-cubic interpolation capability can be used instead of the usual linear interpolation of the magnetic equilibrium that is present in the major grid generators. This last point represents a significant improvement of the accuracy of the magnetic equilibrium as well as the geometry of the cells around the X-points.
}

\section{Future extensions of Gingred}
\label{toc.Future}
The first public release of the code is now available. However, the development of existing or new capabilities are still in process. (i) As examples, the automatic generation of a grid, without the user interactivity, is developed by using an IDL or ascii input file fixing all required boundary conditions and refinement options, but the automatic detection of the magnetic configuration (single-null, double-null or snowflake) from the magnetic equilibrium input file is under investigation. Globally the graphical user interface can be enhanced allowing multiple choices for the user. (ii) Another improvement under development is the inclusion of Gingred in a grid generator module of the python integrated modeling framework OMFIT~\cite{bib_Meneghini_2013_PFR_8,bib_Meneghini_2015_NF_IAEA} that allows interactions between multiple codes. All figures shown here are obtained with python scripts in OMFIT. (iii) Moreover, the algorithm which tracks radial and poloidal lines across and along the poloidal magnetic flux surfaces could be enhanced by using an external ODE integrator. This improvement would reduce the time computation during the refinement and the plate geometry constraints, even if it is not significantly large (within a few minutes for high resolution grids). (iv) In the same topic, other options managing the distribution of poloidal lengths of the cells in the legs during the plate geometry constraint can be developed. The two available options, namely the ratio and the equal poloidal lengths, seem nevertheless robust enough. A successful test has been done for the grid construction of a small-angle-slot (SAS) divertor similar to the one recently installed at DIII-D, and UEDGE convergences have already been obtained. (v) Finally, the generalization of the code is straightforward for an arbitrary number of X-points, even without X-point in the case of field reversed configurations (already developed but not shown here), or with 3 X-points if one wants to consider a cloverleaf divertor, or associate a lower snowflake with a upper single-null configuration.

\section*{Acknowledgements}
This work was performed under the auspices of the U.S. Department of Energy by Lawrence Livermore National Laboratory under Contract DE-AC52-07NA27344. The authors would like to acknowledge M. E. Rensink and T. D. Rognlien for fruitful discussions, and V. A. Soukhanovskii for providing data of simulated NSTX-U magnetic equilibria.\\

\appendix

\section{Topology of main available grids}
\label{toc.Appendix}

The simplest magnetic equilibrium contains only one X-point (called a single-null) and two plates. Here we show the initial $(8 \times 4)$ lower single-null (SNL, where the primary X-point is at the bottom of the tokamak). Fig.~\ref{fig.Grids.SNL} shows the magnetic configuration as well as the 2D logic map of the grid with the values of the minimal set of parameters. \\
A second kind of magnetic topology is obtained when we include the secondary X-point at the top of the machine where the primary X-point is a SNL. Fig.~\ref{fig.Grids.SNL_DNL} shows the magnetic configuration and Fig~\ref{fig.Maps.SNL_DNL} the 2D logic map of the initial $(12 \times 5)$ grid with the values of the minimal set of parameters.

\begin{figure}[!htbp]
\centerline{
$\begin{array}{cc}
\includegraphics[width=40mm]{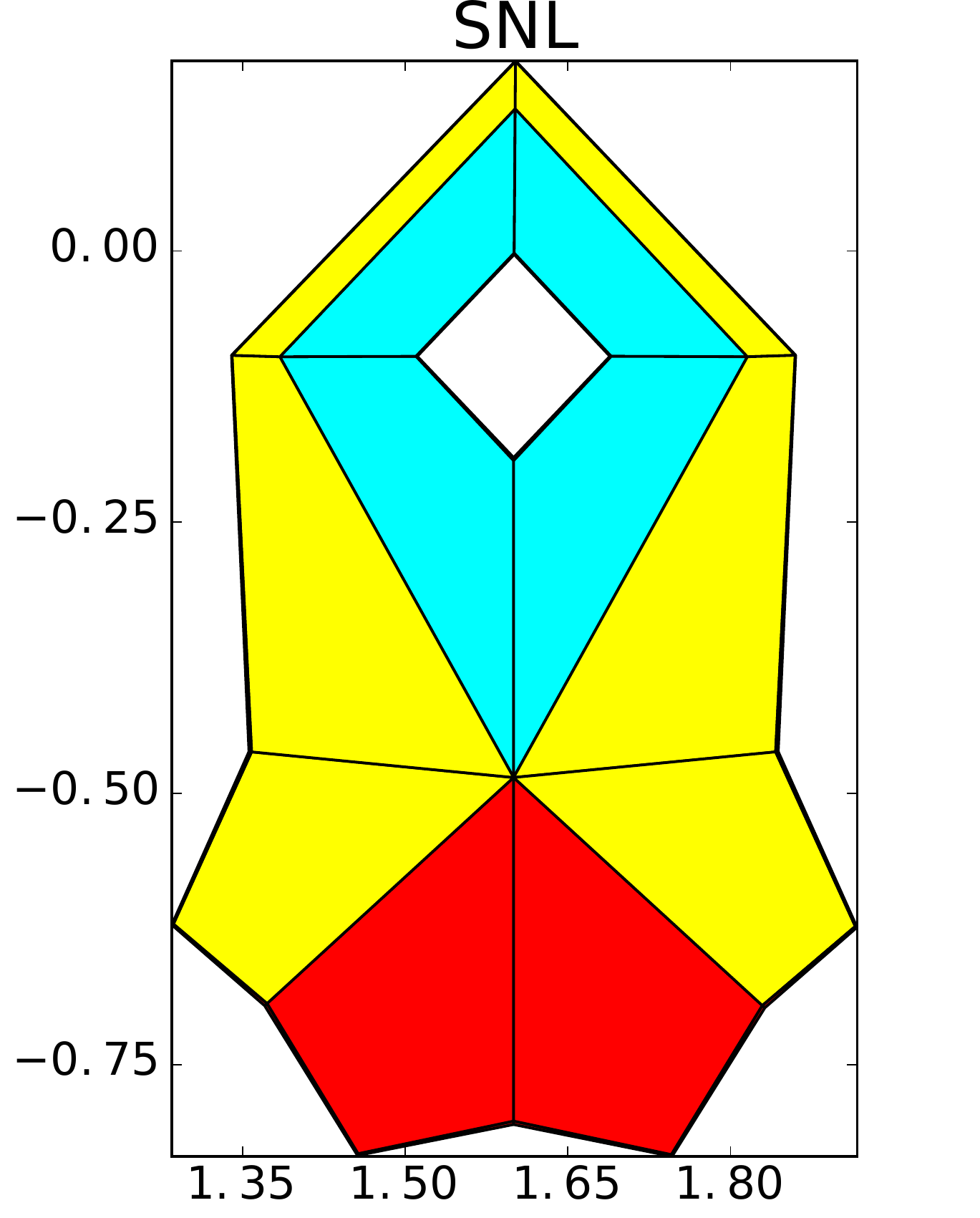} %70mm % 56
&
\qquad \includegraphics[width=28mm]{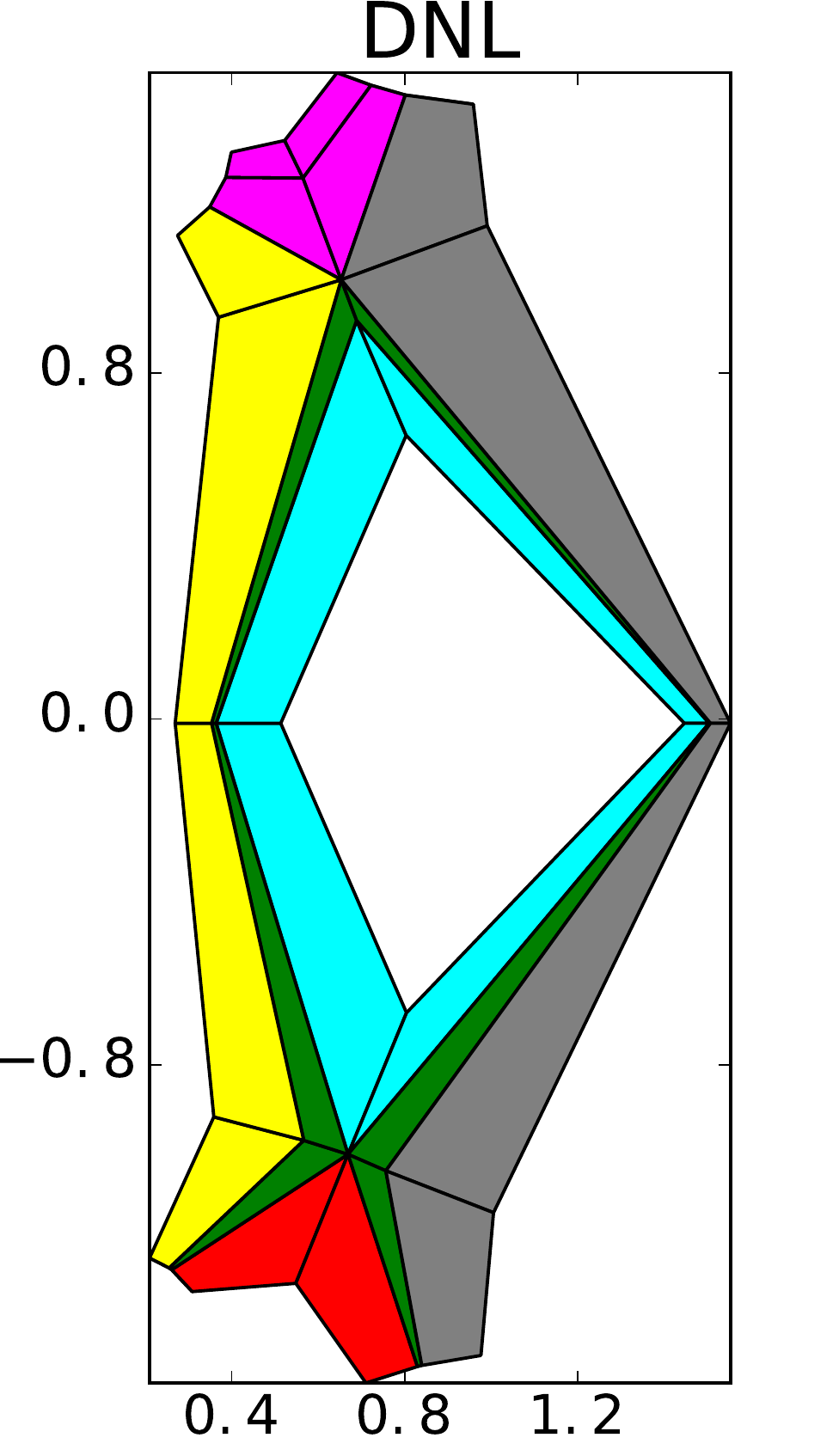} \qquad %70mm %40
\\
\label{fig.Grids.SNL} \textrm{(a) SNL grid}
&
\label{fig.Grids.DNL} \textrm{(b) DNL grid}
\end{array}
$}
\caption{Main single-null and double-null grids}
\label{fig.Grids.SNL_DNL}
\end{figure}
\begin{figure}[!htbp]
\centerline{
$\begin{array}{cc}
\includegraphics[width=70mm]{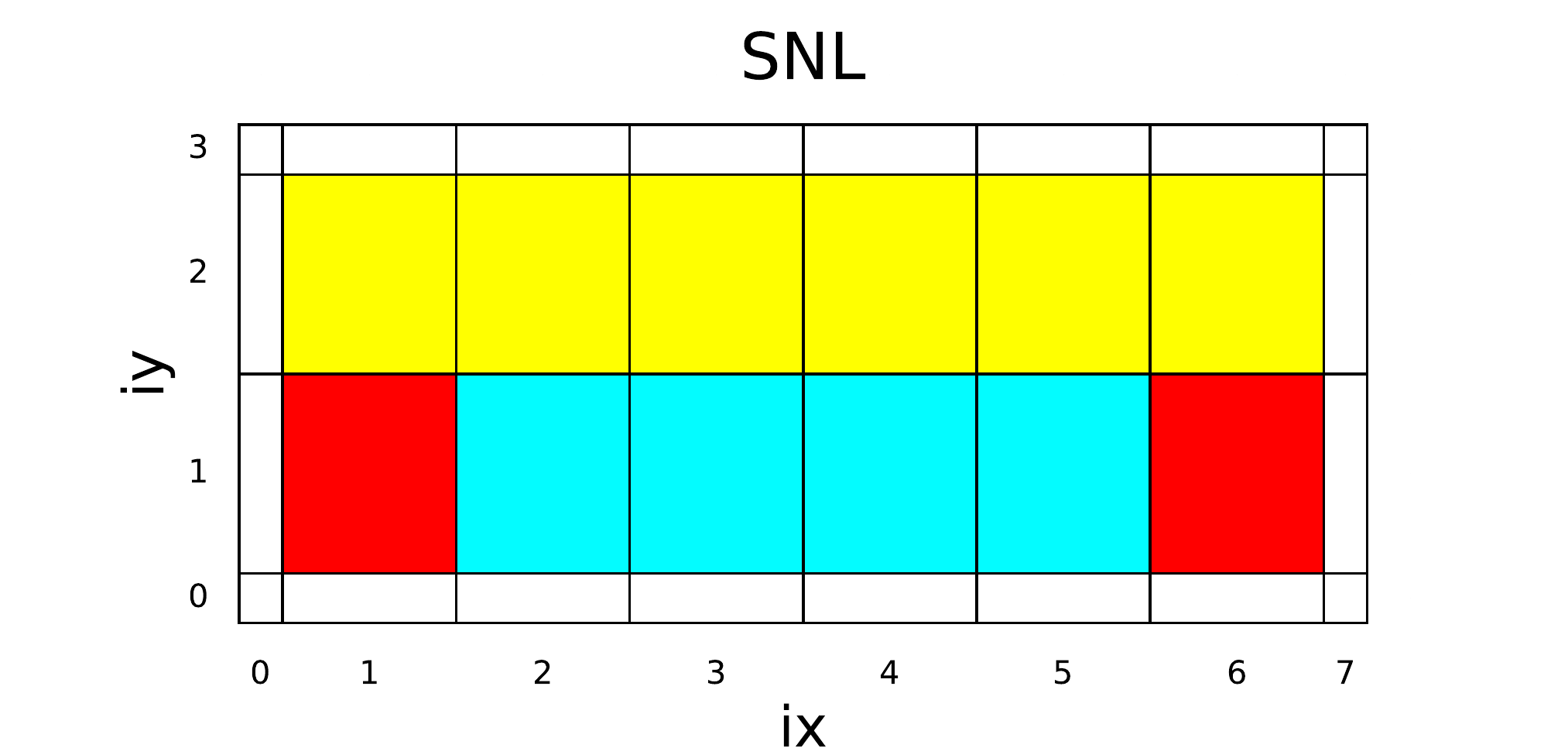} % 55
&
\includegraphics[width=70mm]{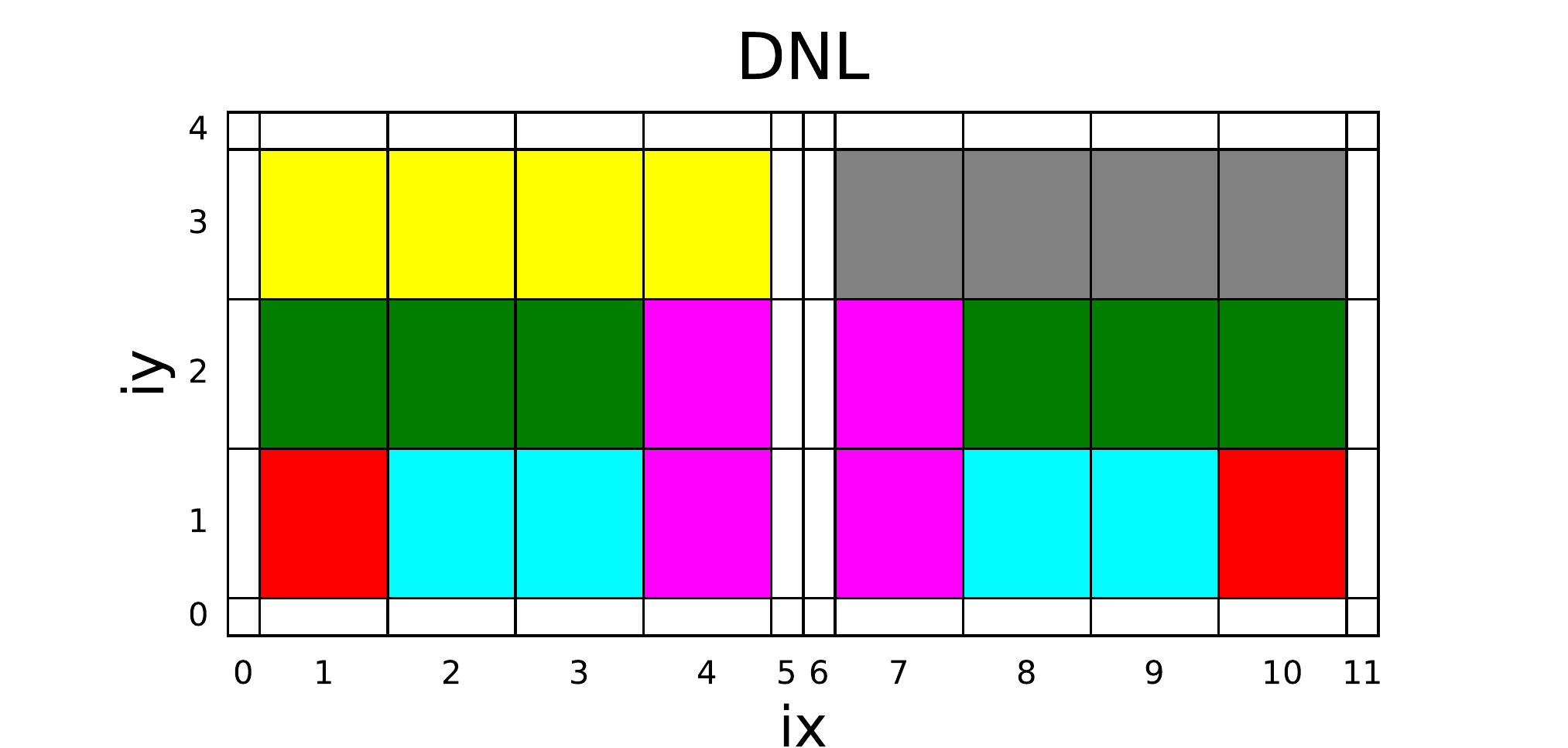} %75mm % 55
\\
\label{fig.Maps.SNL} \textrm{(a) SNL map}
&
\label{fig.Maps.DNL} \textrm{(b) DNL map}
\end{array}
$}
\caption{Main single-null and double-null logic maps. For the single-null, the initial grids are $8 \times 4$ and the position of the X-points and the plates are \cmd{ixpt1[0]=1} ,\cmd{ixpt2[0]=5}, \cmd{ixlb[0]=0} and \cmd{ixrb[0]=6}. For the double-null, the initial grids are $12 \times 5$ and the position of the X-points and the plates are given by \cmd{ixpt1=[1,7]}, \cmd{ixpt2=[3,9]}, \cmd{ixlb=[0,6]} and \cmd{ixrb=[4,10]}.}
\label{fig.Maps.SNL_DNL}
\end{figure}

\updated{The extension of these maps to some snowflake configurations has been developed by Rensink~\cite{bib_Rensink_2017} and is reproduced here.} Snowflake divertor configurations are probably the best candidates to solve the particle and heat exhaust by radiative losses before the first material in the SOL, the divertor plates. It is indispensable to allow UEDGE simulations of snowflake divertors. The starting point is to be able creating snowflake grids. There are two type of snowflake geometries, the snowflake-minus (when the secondary X-point is in the SOL) and the snowflake-plus (when the secondary X-point is in the private flux region of the primary X-point). The two main possible $(13 \times 5)$ grids of snowflake-minus (respectively snowflake-plus) are shown in Fig.~\ref{fig.Grids.SFmp} where the 2D logic maps are shown in Fig.~\ref{fig.Maps.SFmp}.

\begin{figure}[!htbp]
\centerline{
$\begin{array}{cccc}
\includegraphics[width=40mm]{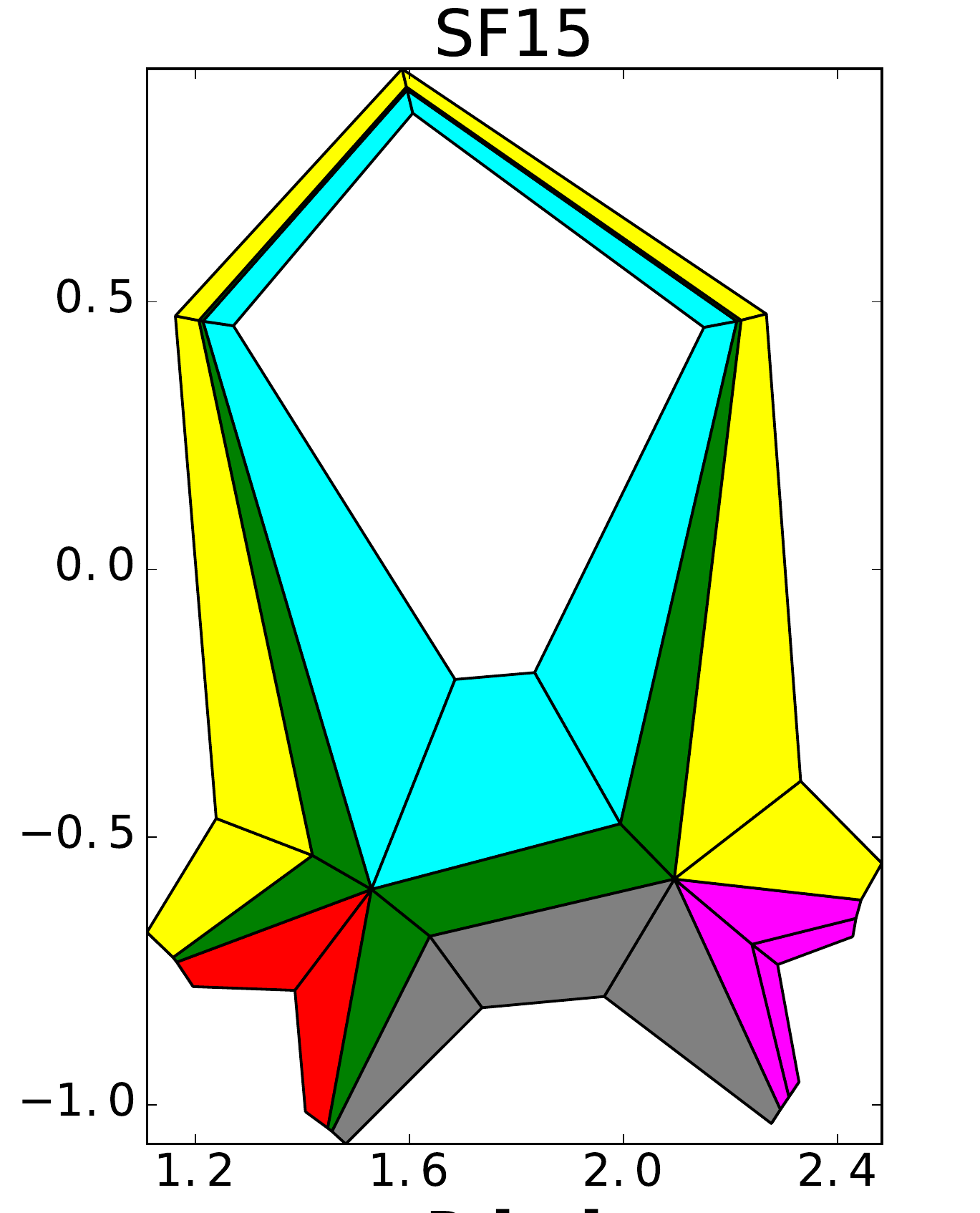} %77mm
&
\includegraphics[width=37mm]{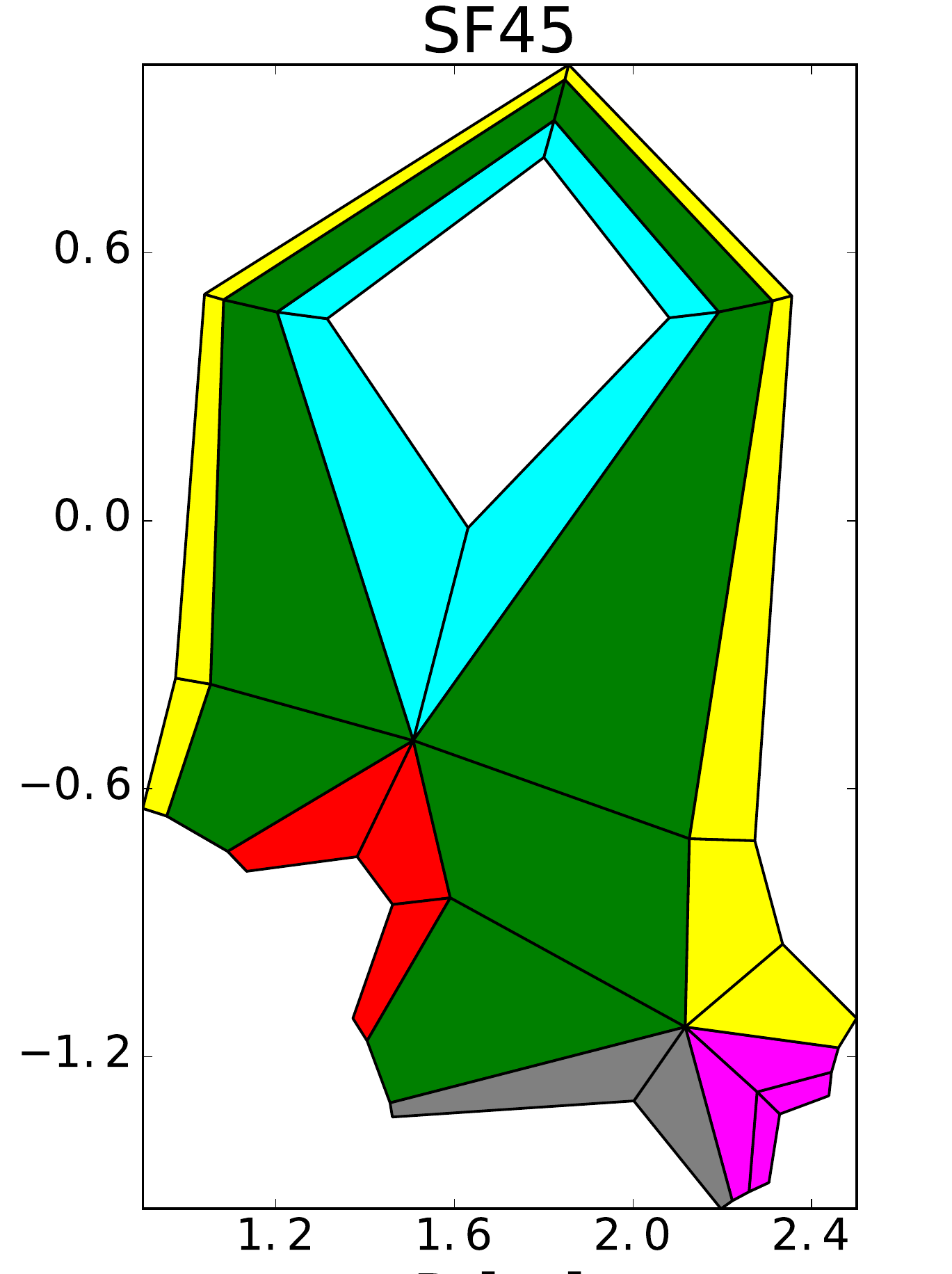} %70mm
&
\includegraphics[width=40mm]{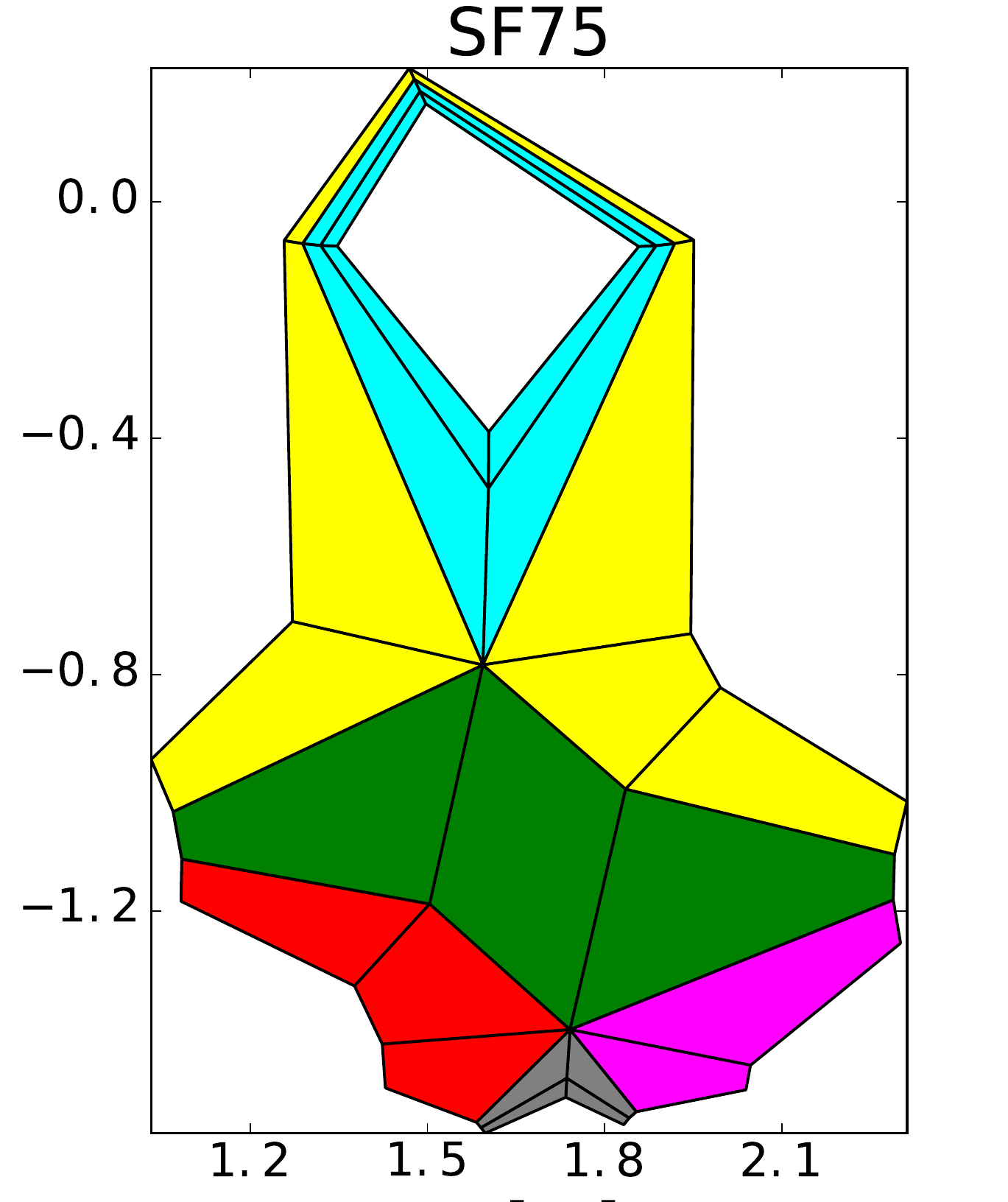} %75mm
&
\includegraphics[width=40mm]{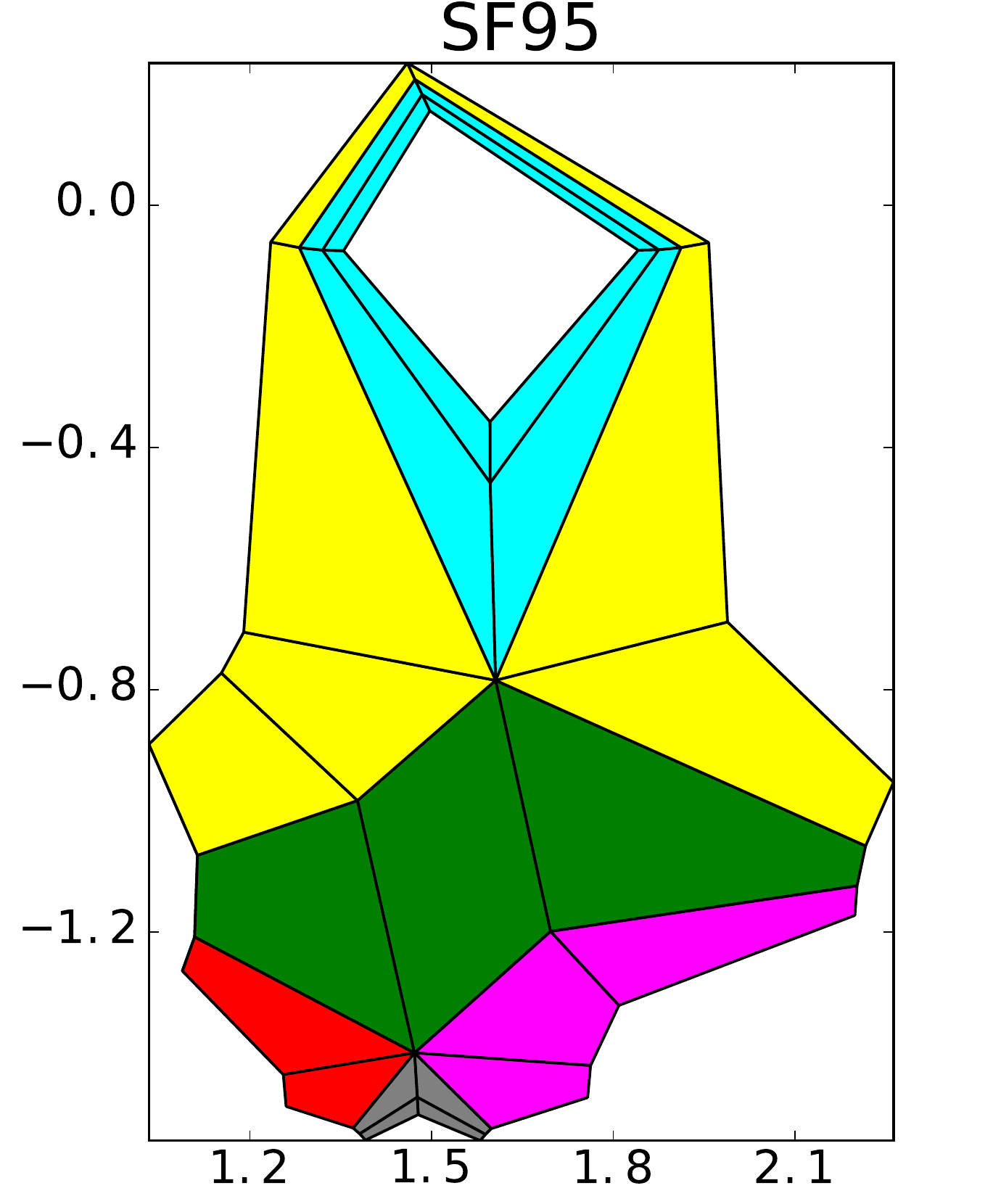} %72mm
\\
\label{fig.Grids.SF15} \textrm{(a) SF15 grid}
&
\label{fig.Grids.SF45} \textrm{(b) SF45 grid}
&
\label{fig.Grids.SF75} \textrm{(c) SF75 grid}
&
\label{fig.Grids.SF95} \textrm{(d) SF95 grid}
\end{array}
$}
\caption{Main snowflake-minus and snowflake-plus grids.}
\label{fig.Grids.SFmp}
\end{figure}
\begin{figure}[!htbp]
\centerline{
$\begin{array}{cc}
\includegraphics[width=70mm]{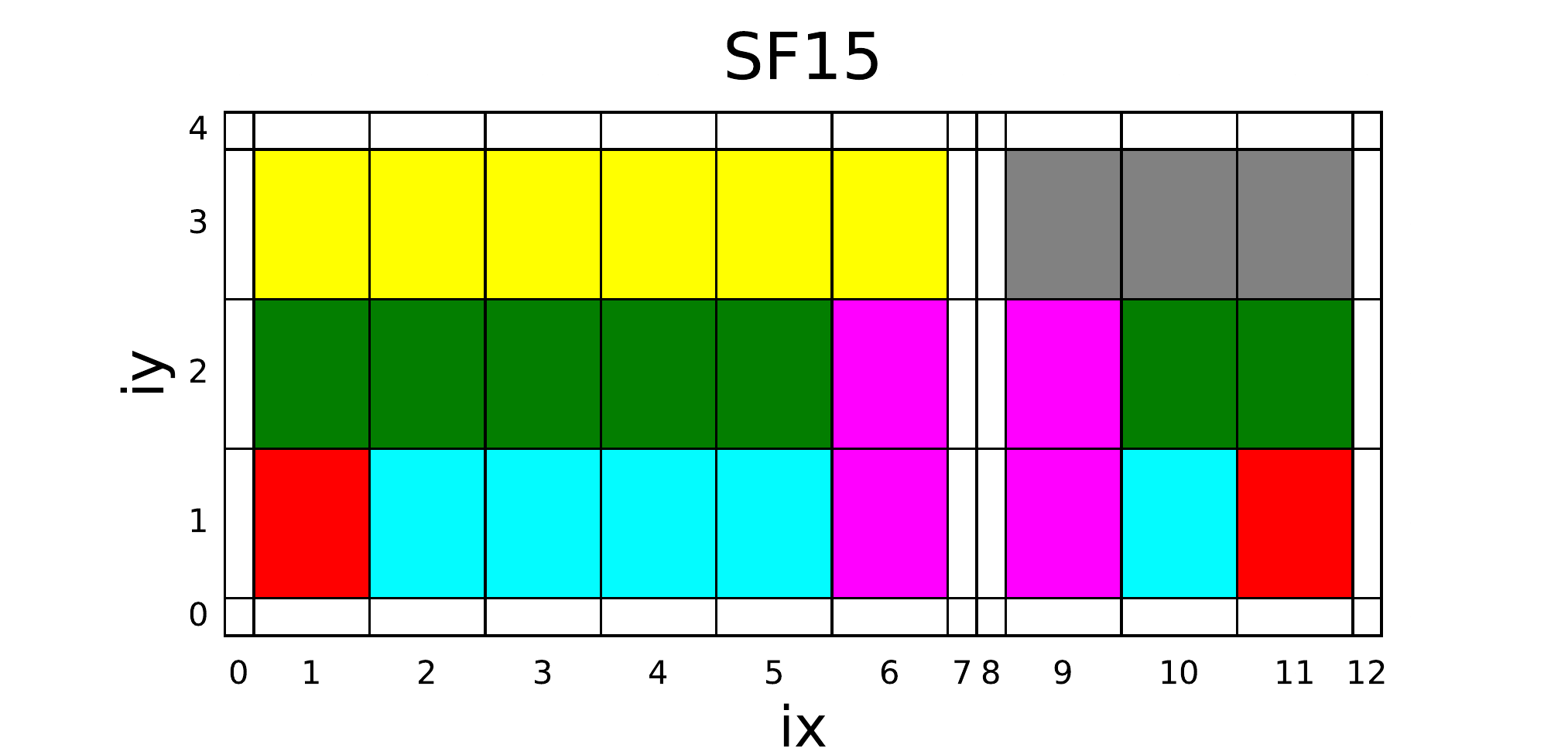} %75mm
&
\includegraphics[width=70mm]{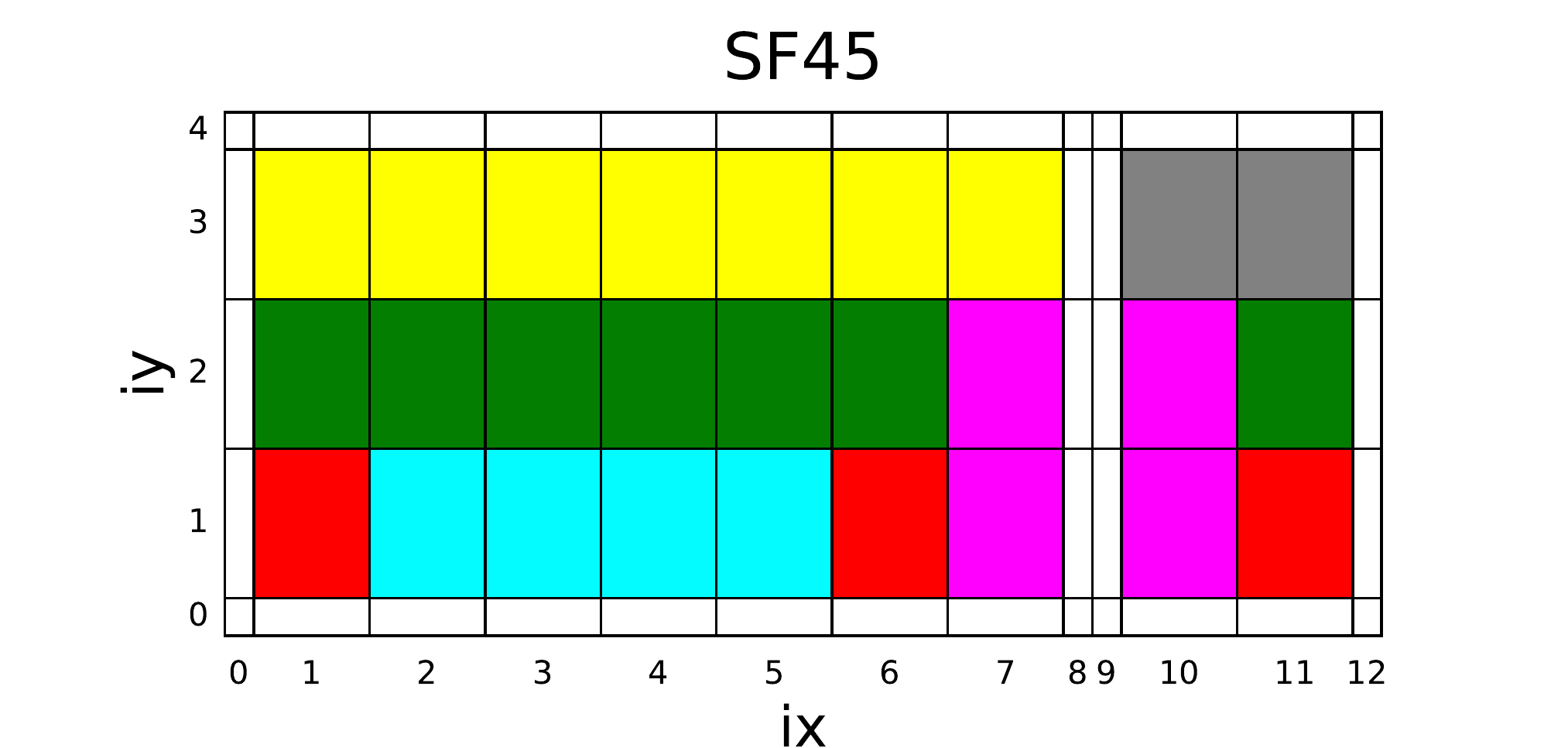} %75mm
\\
\label{fig.Maps.SF15} \textrm{(a) SF15 map}
&
\label{fig.Maps.SF45} \textrm{(b) SF45 map}
\\
\includegraphics[width=70mm]{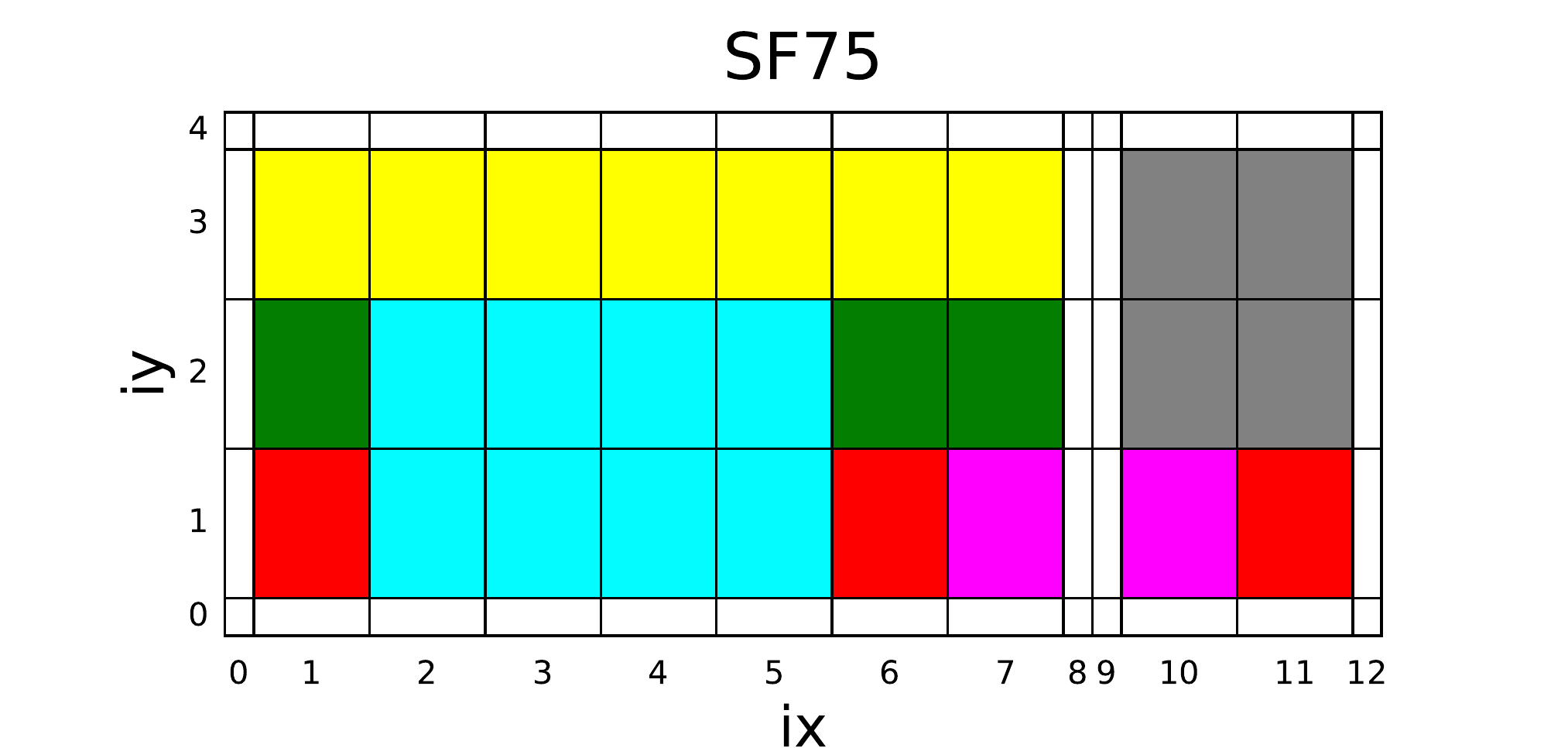} %75mm
&
\includegraphics[width=70mm]{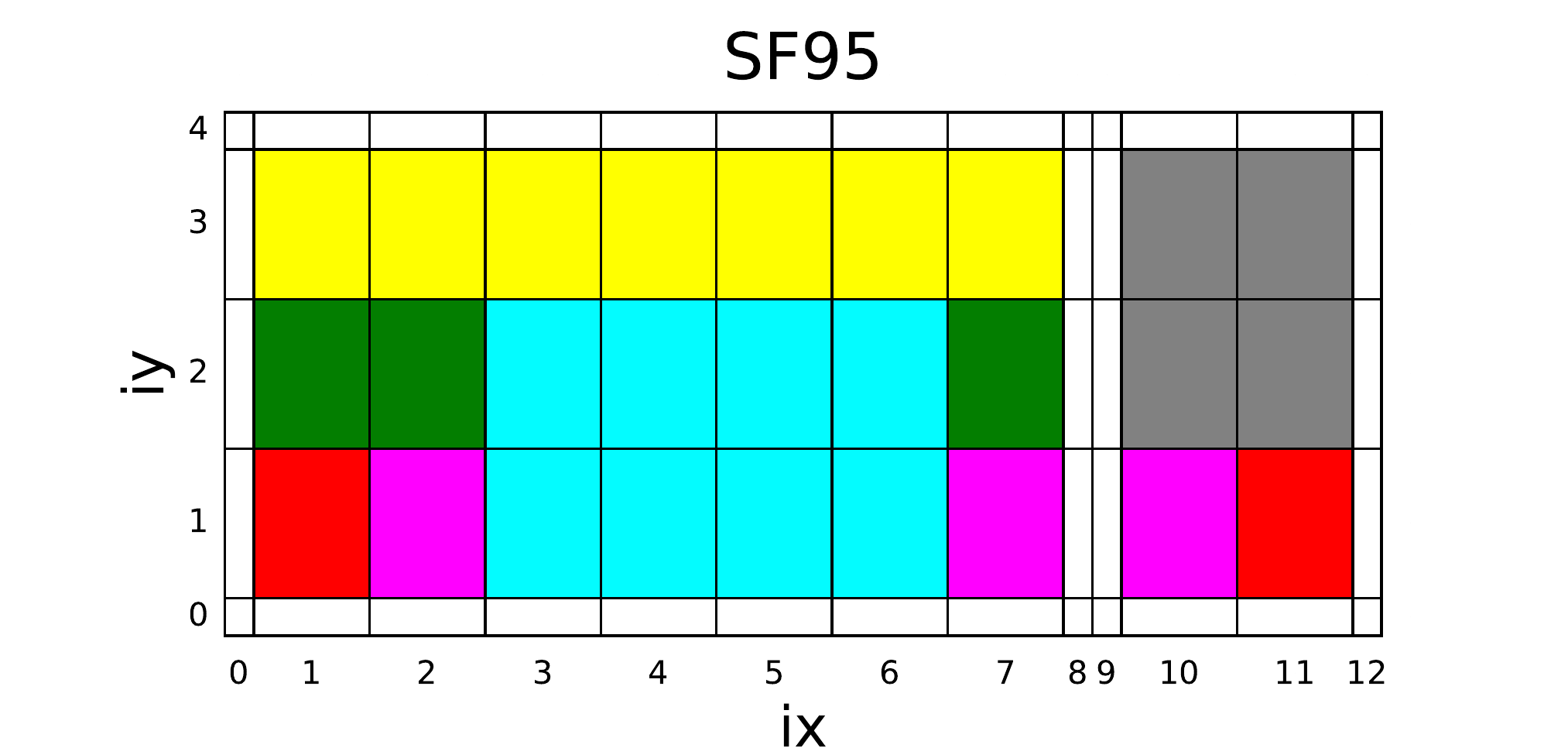} %75mm
\\
\label{fig.Maps.SF75} \textrm{(c) SF75 map}
&
\label{fig.Maps.SF95} \textrm{(d) SF95 map}
\end{array}
$}
\caption{For Main snowflake-minus logic maps: The initial grids are $13 \times 5$ and the position of the X-points and the plates are \cmd{ixpt1=[1,9]} and \cmd{ixpt2=[5,10]}, \cmd{ixlb=[0,8]} and \cmd{ixrb=[6,11]} for the SF15, and \cmd{ixpt1=[1,6]}, \cmd{ixpt2=[5,10]}, \cmd{ixlb=[0,9]} and \cmd{ixrb=[7,11]} for the SF45. For main snowflake-plus maps: The initial grids are $13 \times 5$ and the position of the X-points and the plates are \cmd{ixpt1=[1,6]}, \cmd{ixpt2=[5,10]} for the SF75 and \cmd{ixpt1=[1,6]}, \cmd{ixpt2=[2,10]} for the SF95, and \cmd{ixlb=[0,9]} and \cmd{ixrb=[7,11]} for both.}
\label{fig.Maps.SFmp}
\end{figure}

All indices and notations used in Gingred are given in the caption of figures. All snowflake grids contains $13 \times 5$ cells, including the guard cells. The difference between the 2 snowflake-minus and the 2 snowflake-plus are the poloidal position of the secondary X-point. We can see that the projection of the secondary X-point crossing perpendicularly the flux surfaces in the direction of the primary X-point can be closer to the inner plate P1 (at \cmd{ix=0}) or the outer plate P2 (at \cmd{ix=7} or \cmd{8}). Indeed, for the snowflake-minus grids, the projection of the secondary X-point goes toward the core in cyan (SF15) or primary private flux region in red (SF45). Similarly for the snowflake-plus grids, the projection of the secondary X-point goes toward the low field side SOL (SF75) or the high field side SOL (SF95).

%\vspace{8cm}



%\section*{References}

\end{document}